\documentstyle[11pt,newpasp,twoside,epsf]{article}
\def\edcomment#1{\iffalse\marginpar{\raggedright\sl#1\/}\else\relax\fi}
\reversemarginpar

\begin{document}
\title{Bisector analysis as a diagnostic of intrinsic radial-velocity variations}
 \author{N.C. Santos, M. Mayor, D. Naef, D. Queloz, S. Udry}
\affil{Observatoire de Gen\`eve, CH-1290 Sauverny, Switzerland}

\begin{abstract}
In this contribution we present the results of the application
of the bisector of the cross-correlation function as a diagnostic of 
activity-related radial-velocity variations. 
The results show that the technique is very effective. We present
examples for which the application of the bisector analysis was essential to 
establish the planetary nature of the candidate or to exclude an 
orbital signature. 
An analysis of the behaviour of the bisector for active dwarfs of different spectral types shows that the relation between the bisector and the radial-velocity variation depends in a great extent on the
$v\,\sin{i}$ of the star. The results may shed a new light on the intrinsic sources of radial-velocity variation for different types of solar-type dwarfs.

\end{abstract}

\section{Introduction}

Activity-related phenomena (e.g. spots or convective inhomogeneities) 
can induce radial-velocity variations (Saar \& Donahue 1997; Santos et al. 2000b). These can be very important for high-precision radial-velocity planet-search programmes: the resulting radial-velocity signal can mimic a 
planetary orbital signature and ``produce'' false planetary candidates,
as a result of line shape variations (e.g. bisector changes due to a rotating spot).

Here we describe the use of the bisector of the cross correlation function 
(as described in detail in Queloz et al. 2000) as a diagnostic of activity induced radial-velocity variations. Some examples are shown and the main results are presented and discussed.

\section{Activity vs planetary signature?}

Amongst the stars in the Geneva planet-search programmes\footnote{Both ELODIE and CORALIE; see http://obswww.unige.ch/$\sim$udry/planet/planet.html}, some
have been found to exhibit periodic radial-velocity changes correlated to
bisector variations.  
The standard case is HD\,166435 (Queloz et al. 2000). HD\,166435 is a G0 active dwarf which was found to have a 
periodic radial-velocity signal with a 3.8-day period and 83 m\,s$^{-1}$ of radial-velocity semi-amplitude. 
Although at first the radial-velocity signal seemed to
be connected to the existence of a planetary companion, the same period was observed in photometry (see Queloz et al. 2000, for further details). This 
lead us to study the bisector of the cross-correlation function for 
possible variations. 
The results showed that the bisector inverse slope (BIS) was changing in phase with the radial-velocity signal, proving the intrinsic source of the radial-velocity
variations (Fig.~1, right).

\begin{figure}[t]
\plotone{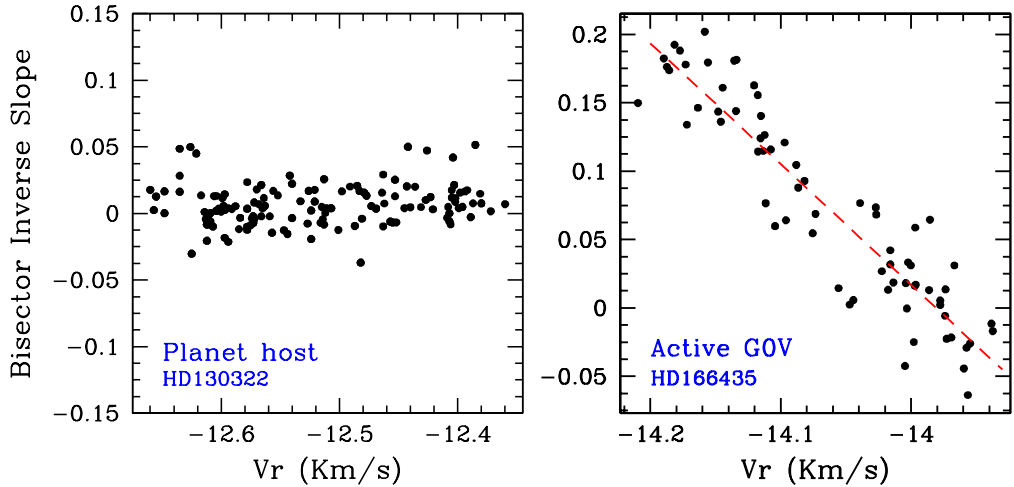}
\caption{Bisector inverse slope as a function of the radial-velocity for two
stars. {\it Left:} HD\,130322, an active K dwarf with a planetary companion; 
{\it right:} HD\,166435 (see text)}
\end{figure}

On the other hand, this technique has been successfully applied to confirm the presence of planetary companions to the solar-type active dwarfs HD\,192263 (Santos et al. 2000a), HD\,130322 (Udry et al. 2000) and HD\,1237 (Naef et al. 2000a, 2000b), all K dwarfs. The case of HD\,130322 can be seen in Fig.~1 (left). No trend is observed, confirming that the radial-velocity signal is not related to changes in the bisector of the cross-correlation function.

\section{Active stars and bisector behaviour}

We have analyzed a sample of 14 active G and K dwarfs 
(defined here as having $\log{R'_{HK}}>-4.4$) from the CORALIE 
and ELODIE planet-search samples, having more than $\sim$20
high-precision radial-velocity measurements. For those stars we have computed the bisector inverse slope to be compared to the radial-velocity values.

To test a possible relation with spectral type we have plotted
in Fig.~2 (left) the values of the slope of the relation BIS vs. RV against the 
colour index $B-V$. 
The observed trend seems to suggest that although active late F and early 
G dwarfs do induce radial-velocity changes by altering the line
profiles, later type active dwarfs are less sensitive to such line asymmetries. That may explain why this latter objects have, for a given activity level, lower radial-velocity ``jitter'' (Santos et al. 2000b). However, as it is clear from Fig.~2 (right), this effect is in great 
extent controlled by the $v\,\sin{i}$ of the star. 
More data at a given $v\,\sin{i}$ is needed to unveil other possible dependences e.g. with colour or activity level.
Stars with planetary companions that were included
after subtraction of the orbital solution do appear in the main trend.

A better knowledge of the relations of Fig.~2 may became useful in the future, since it may represent a way of disentangling between activity-induced and
orbital radial-velocity variations.

\begin{figure}[t]
\plotone{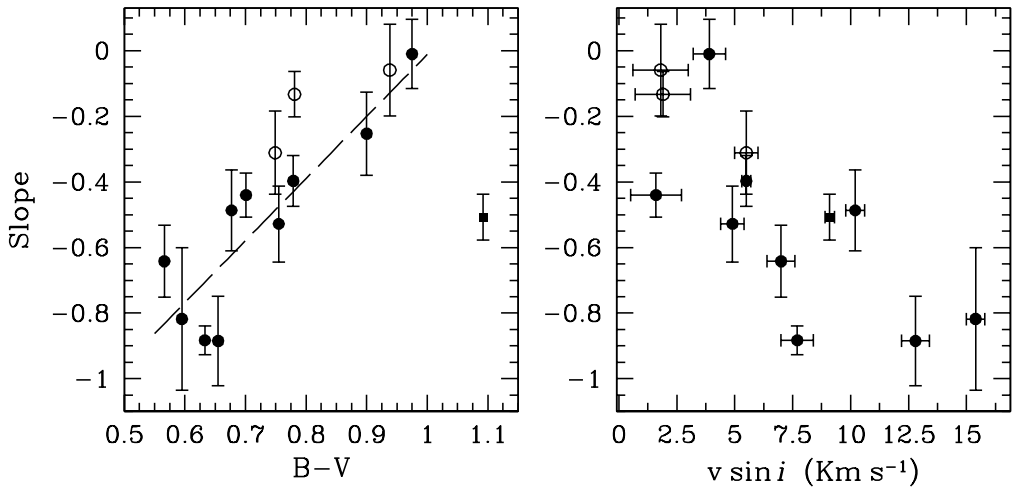}
\caption{{\it Left:} Slope of the relation BIS vs. RV plotted against $B-V$. The best linear non-weighted fit to the points with $(B-V)~<~1$ is presented. Active stars with planets (open dots) are plotted after subtraction of the orbital signal. The square corresponds to a K3 dwarf with a particularly high chromospheric activity index (HD\,29697, $\log{R'_{HK}}\sim-4.1$; Duncan et al. 1991) and $v\,\sin{i}$; it is not included 
in the fit. {\it Right:} Values of the same Slope against $v\,\sin{i}$}
\end{figure}

\end{document}